\documentclass[aps, pre, twocolumn, superscriptaddress]{revtex4-1}

\usepackage{mathrsfs}
\usepackage{amsfonts}           
\usepackage{amssymb}            
\usepackage{amsmath}            
\usepackage{comment}
\usepackage{latexsym}           
\usepackage{dcolumn}            
\usepackage[dvips]{graphicx}   
\usepackage{enumerate}
\usepackage{epstopdf}           
\usepackage{fancyhdr}           
\usepackage{hyperref}           
\usepackage{setspace}           
\usepackage{xspace}             
\usepackage{subfigure}          
\usepackage{color}   
\usepackage{bm}                 
\usepackage[ulem=normalem]{changes}
\usepackage{multirow}
\usepackage{adjustbox}
\newcommand{\bfI}{\bm{I}}
\newcommand{\phij}{\phi_{j}^{\epsilon}}
\newcommand{\veps}{\varepsilon}

\newcommand{\Cspace}{\mathbb{C}}
\newcommand{\Xdag}{{\bf X}^\dagger}
\newcommand{\bfw}{{\bf w}}

\newcommand{\bfPsi}{{\bf\Psi }}
\newcommand{\bfb}{{\bf b}}
\newcommand{\bfphi}{{\bf\phi}}
\newcommand{\dav}{{\langle d\rangle}}
\newcommand{\sav}{{\langle \sigma_{zz}\rangle}}
\newcommand{\libPsi}{{\rm lib}_{\Psi}}
\newcommand{\bfJ}{{\bm{J}}}

\newcommand{\bfU}{{\bf U}}
\newcommand{\bfV}{{\bf V}}
\newcommand{\bfSg}{{\bf\Sigma}}
\newcommand{\bfVstar}{{\bf V}^*}
\newcommand{\bfX}{{\bf X}}
\newcommand{\bfY}{{\bf Y}}

\newcommand{\si}{\sigma}

\newcommand{\bea}{\begin{eqnarray}}
\newcommand{\eea}{\end{eqnarray}}

\newcommand{\cf}{{\it cf.}~}

\newcommand{\eps}{\epsilon}

\newcommand{\eg}{{\it eg.~}}
\newcommand{\ie}{{\it ie.\,}}

\newcommand{\bfr}{{\bf r}}

\begin{document}

\title{Microstructural Inelastic Fingerprints And Data-Rich Predictions of Plasticity and Damage in Solids}
\date{\today}
\author{Stefanos Papanikolaou}
\affiliation{Department of Mechanical and Aerospace Engineering, West Virginia University, Morgantown, WV26506, United States.}
\affiliation{Department of Physics, West Virginia University, Morgantown, WV26506, United States.}


\begin{abstract}
Inelastic mechanical responses in solids, such as plasticity, damage and crack initiation, are typically modeled in constitutive ways that display microstructural and loading dependence. Nevertheless, {linear} elasticity at infinitesimal deformations is used for microstructural properties. We demonstrate a framework that builds on sequences of microstructural images to develop  fingerprints of inelastic tendencies, and then use them for data-rich predictions of  mechanical responses up to failure. In analogy to common fingerprints, we show that these two-dimensional instability-precursor signatures may be used to reconstruct the full mechanical response of unknown sample microstructures; this feat is achieved by reconstructing appropriate average behaviors with the assistance of a deep convolutional neural network that is fine-tuned for image recognition. We demonstrate basic aspects of microstructural fingerprinting in a toy model of dislocation plasticity and then, we illustrate the method's scalability and robustness in phase field simulations of model binary alloys under mode-I fracture loading.
\end{abstract}
\maketitle 

\subsection{Introduction}

{
A critical bottleneck in the systematic material discovery, optimization and deployment is the lack of consistent and robust microstructure-property relationships that hold across environmental conditions and material classes. A major reason for this deficiency  lies in the fact that the inelastic mechanical response of solids does not always originate in visible defects of the microstructure, but requires additional dynamical insights, not always accessible or/and comprehensible~\cite{Raabe:1998aa}.

Consequently, available modeling approaches for mechanical applications, such as discrete or continuum models of molecular or continuum microstructural dynamics~\cite{Kanoute:2009yl,groh2009advances,Raabe:1998aa} require robust and multi-scale physical understanding and associated constitutive laws, with applicability in technology-relevant extreme conditions, such as high temperatures, pressures and strain-rates. 
In contrast, in a very wide range of mechanical applications in solids at infinitesimal deformations (\ie less than 2\% strain), elastic moduli are found to be consistent and robust, defining critical aspects of material behavior~ \cite{Ostoja-Starzewski:2007aa,Ashby:1993aa}. Moreover, the so-called deformation superposition principle~\cite{Asaro:2006fr}, stating that elastic deformations may be superposed to deformations of other origin (plastic, damage), has led to the development of consistent defect theories and methods (\eg dislocation mechanics)~\cite{bulatov2006computer,Raabe:1998aa}.

\begin{figure*}[tbh]
\centering
\includegraphics[width=0.95\textwidth]{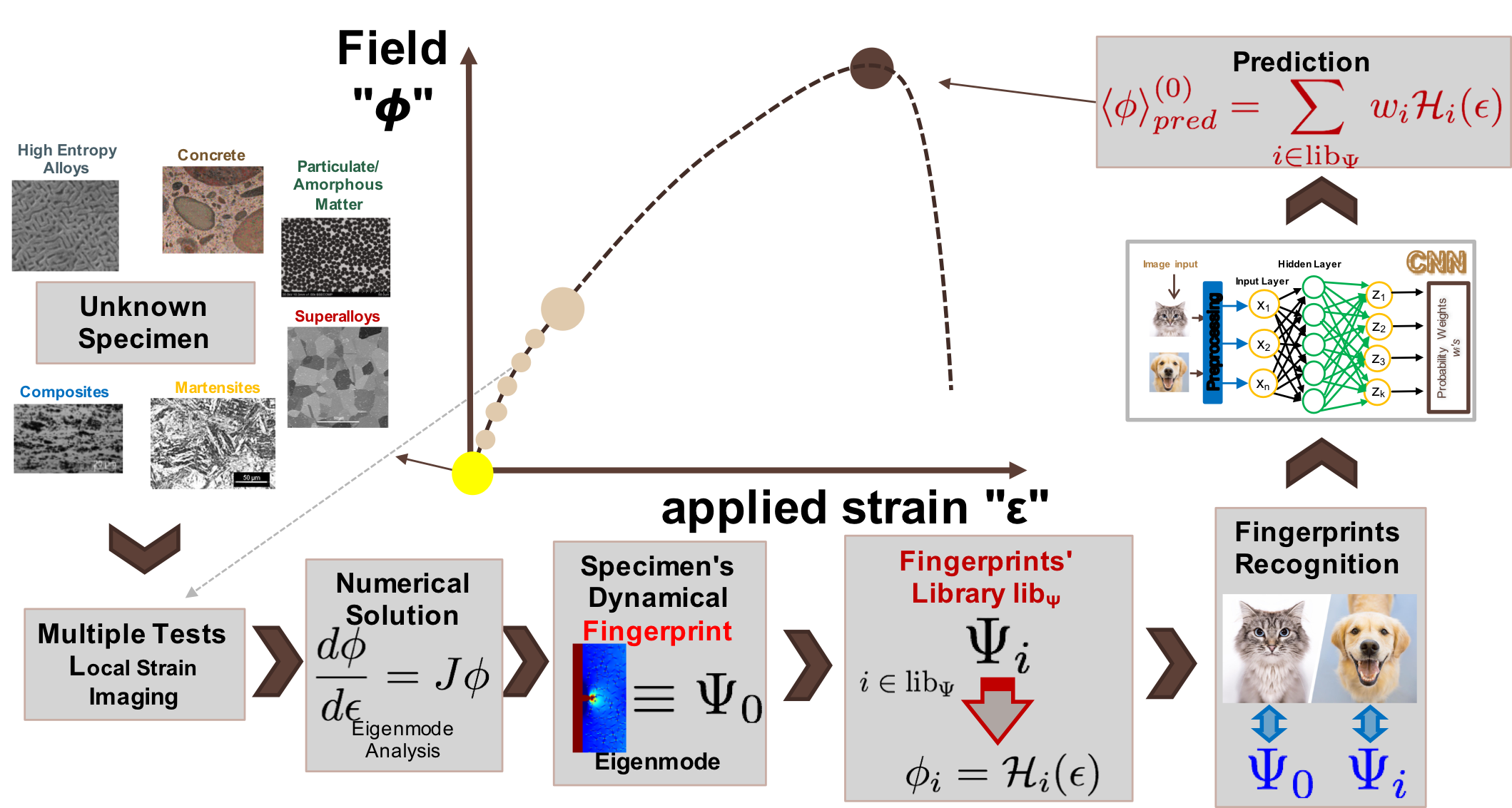}
\caption{{\bf Process diagram of main steps towards equation-free prediction for plasticity and damage.} The field $\phi$ typically is a strain field, invariant under frame rotations, that can be accurately measured through image correlation techniques. The applied strain $\epsilon$ corresponds to the externally-controlled loading steps.}
\label{fig:plan}
\end{figure*}

In this paper, we propose that small-deformation elastic strain image sequences (up to 2$\%$ strain), readily produced experimentally~\cite{Schreier:2009xd}, may be used as input towards calculating ``microstructural fingerprints” in the form of dynamical modes of inelasticity, which we label as Elastic Instability Modes (EIM). EIMs are produced by analyzing surface strain image sequences on a loaded sample, and consist of physically predominant perturbations to the elastic mechanical response. EIMs may be considered as a dimensional hyper-reduction of the surface strain image sequence towards a characteristic image that captures the surface strain evolution under loading. We develop the theory for the calculation of EIMs from image sequences, and we demonstrate it for a toy model of dislocation plasticity, as well as a realistic phase-field model of damage and fracture. Finally, we demonstrate that the combination of these microstructural fingerprints with deep convolutional neural networks (that can be used to classify and distinguish EIMs) can lead to detailed mechanical response predictions for unknown microstructures and samples. In this way, we display a robust protocol for the characterization and understanding of unknown microstructures, using only inexpensive, surface data at small deformations.
The overall approach is labeled as ``Stability of Elasticity Analysis" (SEA) (see Fig.1)  and may complement available multiscale microstructural modeling approaches~\cite{Raabe:1998aa}. 

The role of SEA is two-fold: First, to characterize in a concrete way the progression of local mechanical instabilities in a material by the understanding of EIMs in experiments and modeling. Second, to store EIMs as classification tools in a data library, labeled as $\libPsi$, together with any tested mechanical responses of interest (\eg uniaxial loading data at 30\% strain), and then use them as microstructural ``fingerprints" for producing mechanical predictions by reconstructing them through deep convolutional neural networks (dCNN)~\cite{dnn}. The use of dCNNs is not SEA's requirement, but may be ideal for physically permissible input data superpositions. 

The usefulness of SEA lies in the development of material design strategies~\cite{Ashby:1993aa}. 
In the large multidimensional parameter space of possible compositions and loading conditions, a natural bottleneck has been the consistent prediction of mechanical behaviors of new compounds by testing single-dimensional parameter lines. A multitude of data science and machine learning~\cite{Liu:2015hp,Liu:2017ii,Ramprasad:2017nw,Zhang:2018ez,pap2019,Papanikolaou:2019aa,Papanikolaou:2019ab} approaches have been recently proposed for materials and mechanics~\cite{Kirchdoerfer:2016pv}. In particular, image recognition methods have been efficiently used for the identification of ``flaws" in materials and structures~\cite{def1,def2,def3,def4,Cubuk:2015aa}, and work remarkably well  when the flaw's mechanical effects are well understood~\cite{Gobel:2009vn,Salmenjoki:2018aa,Steinberger:2019aa}. In this context, SEA aims to complement these efforts by adding additional physical insights and understanding into promoting studies and predictions with direct experimental connections.

The physical key to the dimensional hyper-reduction method presented in this work is the understanding that in many practical situations,  solids can be considered as initially stable elastic media that are then, mechanically loaded. If    the loading scale is perceived as time (\ie assuming that viscous solid effects are negligible~\footnote{
In the context of elasticity, it is natural to view the loading process of a mechanical system as a non-linear dynamical system with ``time" defined by loading. In this definition, the dynamics in-between loading steps is considered fully dissipative, and thus is neglected. 
}), then this problem can be phrased in the context of non-linear dynamical systems~\cite{guck}: A loaded solid can be thought of near a stable fixed point, that of elasticity. In general, an $N$-dimensional non-linear dynamical system  consists of a list of variables $x_i(t)\;i=1,2,\cdots,N$, their dynamics ${\dot {x}}_{i}=f_{i}(x)$, and their initial conditions $x_{i}(0)=x_0=0$. In such systems, the characterization of a fixed point (\ie $f_i(x_0)=0$) is controlled by the growth rates of generic perturbations which are given by the spectrum of Lyapunov exponents (LE)~\cite{lyap1,lyap2,lyap3,lyap4}
$\{\lambda _{1},\lambda _{2},\ldots ,\lambda _{n}\}\,$. LEs are the real parts of the eigenvalues of the $N\times N$ Jacobian matrix $J\equiv J_{ij}(t)=\left.{\frac {df_{i}(x)}{dx_{j}}}\right|_{x(t)}$, which provides the fixed-point dynamical evolution $ {\dot {X}}=J X$  where $X=\{x_i\}$. Away from the stable elasticity fixed point, a solid may be unstable to necking,  buckling, plasticity, crack initiation, damage~\cite{Asaro:2006fr}, and while these instabilities do not need to be elastic in origin, they always do have an elastic footprint which may be identifiable at small loads. It is useful to consider these instabilities of elasticity as spatially dependent bifurcations~\cite{Aifantis:1992gh,Bigoni:2012ua,Papanikolaou:2017zl} that could be captured by identifying the system's predominant LEs.

In practical situations, LEs' calculation is numerical, with various available approaches~\cite{Benettin:1980jo,Benettin:1980mw,Brown:1991ay,Bryant:1990ss,Bryant:1993fc,Eckmann:1985yo,Miller:1964uf,Schmid:2010bd,Shimada:1979qc}. However, it has been difficult to achieve significant accuracy with limited experimental data. In this work, we propose a method that is efficient and focuses on the predominant LEs of this loading evolution. In what follows, we present the general framework of SEA in Section~\ref{sec:B}, and then we demonstrate how it works in an exactly solvable toy example of edge dislocation gliding and nucleation in Section~\ref{sec:C}. Then, in Section~\ref{sec:D}, we explore its applicability in a model binary alloy that is simulated using phase-field modeling for elasticity, plasticity and damage in a quite realistic scenario. Finally, in Section~\ref{sec:E}, we demonstrate how to use dCNNs for mechanical predictions of fracture, based on the classification and understanding of microstructural fingerprint EIMs. We conclude with a discussion of future plans and modeling/experimentation applications in Section~\ref{sec:F}.

\subsection{General framework of microstructural fingerprinting based on inelastic signatures}
\label{sec:B}

The core principle behind this work lies in the ability to characterize defected microstructures in terms of the elastic fields generated when small, consecutive loads are applied on them. This principle has been traditionally used towards qualitative insights and understanding of mechanical failure in materials. Possible examples could be the insightful calculation of size effects in notched and disordered specimens~\cite{Bazant:1997hx} \cite{Alava:2006vn} or the elastic fields around dislocations~\cite{Hirth82} that may influence various mechanical properties~\cite{Mura:2013aa}, especially at small scales~\cite{uchic2009plasticity}.

Here, without loss of generality, we consider a scalar field $\tilde\phi^{\eps}(\bfr)$ at applied strain $\eps(t)$, that will represent an observable field of interest on $N^2$ possible surface locations ($\phij$), assumed to have a spatial resolution down to a characteristic scale that defines a square $N\times N$ grid~\footnote{
Here, we assume a square grid, which could be generalized to any grid in a straightforward manner
}, controlled either by practical means (\eg image/camera resolution) or theoretical ones (\eg interatomic distances). The field $\tilde\phi^{\eps}$ may either be a direct elastic field, such as elastic strain or stress, or a field that is strongly correlated to elastic fields, such as the plastic distortion or damage~\footnote{
A relation of plasticity/damage to elastic fields is derived through the elastic fields generated by corresponding defects (\eg dislocations or voids)}. 
In this paper, we will be focusing on $\tilde\phi$ being either the first strain invariant $I_\eps=\veps_{xx}+\veps_{yy}+\veps_{zz}$ or the damage field $d$, which is defined through the parent material's elasticity properties $d = \sqrt{1-C_{ijkl}/C^{0}_{ijkl}}$ with $C^{0}$ corresponding to the undamaged elastic coefficients~\cite{Roters:2019be}.

Assuming the loading of a sample location through an imposed strain profile $\epsilon(t)$, then $\tilde\phi$'s evolution infinitesimally away from the initially elastic fixed point ought to resemble the equation:
\bea
\frac{d\tilde\Phi}{dt}= \rm{C}
\label{eq:1}
\eea
where $\tilde\Phi=\{\tilde\phij\}$, and $\rm{C}$ represents a function of the elastic coefficients. As loading progresses, it is natural to assume a generic leading-order perturbation $\bfJ\tilde\Phi$ to the right side of Eq.~\ref{eq:1}, with $\bfJ$ a $N^2\times N^2$ matrix. If one also proceeds with subtracting the mean response $\phi\equiv\tilde\phi -\rm{C}t$~\footnote{
The subtraction, in this case, amounts to neglecting the average mechanical response. The subtraction may be done locally, by subtracting the expected elastic solution in the microstructure. The investigation of the local subtraction approach is beyond the purposes of the current work.
}, then one has the normal form:
\bea
\frac{d\Phi}{dt}= \bfJ \Phi\;,
\label{eq:norm}
\eea
where $\bfJ$ is typically an unknown matrix. $\bfJ$ is a matrix that controls the most singular LEs. Analogous considerations may be made for other observables such as damage, stress or strain fields. 

The pursuit in understanding instabilities of elasticity requires the precise identification of $\bfJ$ (see also Fig.~\ref{fig:plan}). $\bfJ$ is understood for exactly solvable cases, such as a dislocation pile-up at a precipitate under shear, an elliptical notch under lateral load, or an Eshelby inclusion in an elastic matrix~\cite{Asaro:2006fr}: It can be directly shown that $\bfJ$ captures the long-range stress {\it changes} during subtle movements of inelastic defects (dislocations/micro-cracks/damage/inclusions).~\cite{Kanoute:2009yl, Asaro:2006fr}. Here, we develop an {automatic} framework (SEA) that calculates instability growth exponents and eigenmodes $\bfPsi$ of Eq.~\ref{eq:norm}.  SEA  identifies {\it elastic instability modes} (EIM) $\bfPsi$ through spatially resolved {\it in-situ} image sequences, which solve the eigen-problem of Eq.~\ref{eq:norm} in a least-squares sense. The set of modes $\bfPsi$ of a sample may be considered as its microstructural {fingerprints} for mechanical behavior. 

The numerical estimation of modes that solve Eq.~\ref{eq:norm} proceeds by identifying a sequence of images at $T$ consecutive equidistant strains $\eps_0,\eps_1,\eps_2,\cdots,\eps_n\cdots\eps_{t}$, with $\Delta\eps\equiv\eps_2-\eps_1$. The testing strain $\eps_{t}$ is assumed to be less than $2\%$ and defines the maximum deformation imposed on a sample for SEA testing and identification purposes. In the case of equidistant strains in the sequence, Eq.~\ref{eq:norm} may be rewritten in its discrete form:
\bea
\Phi_{n+1} = \tilde\bfJ \Phi_{n}
\label{eq:disc}
\eea
with $\tilde\bfJ \equiv \bfJ\Delta\eps + \bfI$.
 Then, the predominant EIMs may be identified by solving Eq.~\ref{eq:disc} for $\tilde\bfJ$ by using the Arnoldi algorithm~\cite{Arnoldi:1951aa}. Field $\phi$  $[T\times N^2]$-dimensional matrices are defined, where rows define time/strain evolution while columns define spatial locations: 
\bea
\bfX=\{\bfphi^{\epsilon_0}_j, \bfphi^{\epsilon_1}_j\cdots \bfphi^{\epsilon_n}_j\cdots\bfphi^{\epsilon_{t-1}}_j\}
\eea 
and 
\bea
\bfY=\{\bfphi^{\epsilon_1}_j, \bfphi^{\epsilon_2}_j\cdots \bfphi^{\epsilon_{n+1}}_j\cdots\bfphi^{\epsilon_{t}}_j\}.
\eea
Then, the optimal solution for $\tilde\bfJ$ is~\cite{Rowley:2009fk,Schmid:2010uq,guck,Arnoldi:1951aa}
\bea
\tilde\bfJ_N = \bfY \Xdag
\eea
where $\Xdag$ is the Moore-Penrose pseudo-inverse of $\bfX$. The numerically identified operator $\tilde\bfJ_N$ is the least-squares/minimum-norm solution to the potentially over or under-constrained problem $\tilde\bfJ \bfX =\bfY$. That is, the choice $\bfJ_N$ minimizes the Frobenius norm $||\tilde\bfJ_N \bfX - \bfY||$. The eigenvectors and eigenvalues of $\tilde\bfJ_N$ can be calculated exactly through an exact diagonalization~\cite{Press:1987oq}. Nevertheless, given the possibility of memory issues, one would be primarily interested only on the most predominant eigenvalues and eigenmodes that correspond to the predominant perturbations to elasticity. The most predominant eigenvalues and eigenmodes may be estimated by taking the reduced Singular Value Decomposition (SVD) of $\bfX$:
\bea
\bfX = \bfU \bfSg \bfVstar
\eea
where $\bfU\in\Cspace^{N^2\times r}$, $\bfSg\in\Cspace^{r\times r}$ and $\bfV\in\Cspace^{T\times r}$ with $r$ the rank of $\bfX$. In this work, we maintain $r=8$. Clearly, the singular value amplitudes $\varsigma_j$ of the SVD modes capture the variability in the microstructural evolution, as it is seen through the spatial correlations of the $\Phi$-field. The eigen-decomposition of $\bfU^*\bfY\bfV\bfSg^{-1}$ can be then exactly calculated, giving a set of eigenvectors $\bfw$ and eigenvalues $\mu$. Then, the operator $\tilde\bfJ$ has eigenvalues $\mu$ and eigenvectors 
\bea
\bfPsi=\frac{1}{\lambda}\bfY\bfV\bfSg^{-1}\bfw
\eea
labeled as EIMs. This low-rank approximation of eigenvalues and eigenvectors of $\bfJ$ allows for the approximate reconstruction of the time evolution as:
\bea
\tilde \bfphi(\eps)=\sum_{k=1}^r b_k(0)\psi_k(\eps)e^{\ln\mu_k \eps/\Delta \eps}
\label{eq:pred}
\eea
with the coefficients $b_k(0)$ characterize the initial condition $\bfphi_0$ and $\bfb=\bfPsi^\dagger\bfphi_0$. The quantity $\lambda_k\equiv\ln\mu_k/\Delta\eps$ has a real part, which if larger than 0 signifies a finite instability growth rate, dominated by mode $k$. An imaginary part signifies additional oscillatory response. 

The resolution scale of the $\phi$-field shall be at the characteristic scale of the elastic fluctuations (\eg at the  scale of a Representative Volume Element (RVE))~\cite{Raabe:1998aa} and thus, the dimensions of $\bfJ$ are necessarily finite. 

The analogy of eigenmodes $\Psi$ to {\it fingerprints} is insightful: The sought experimentally relevant strain deformation data sets~\cite{Schreier:2009xd}are typically two-dimensional (2D) and only capture small  surface strains. However, in the same way that 3D humans are being recognized by 2D fingerprints, it is quite plausible that material microstructures may be recognizable by the load-dependent elastic defect signatures in the small strain regime.  

\begin{figure}[tb]
\centering
\includegraphics[width=0.5\textwidth]{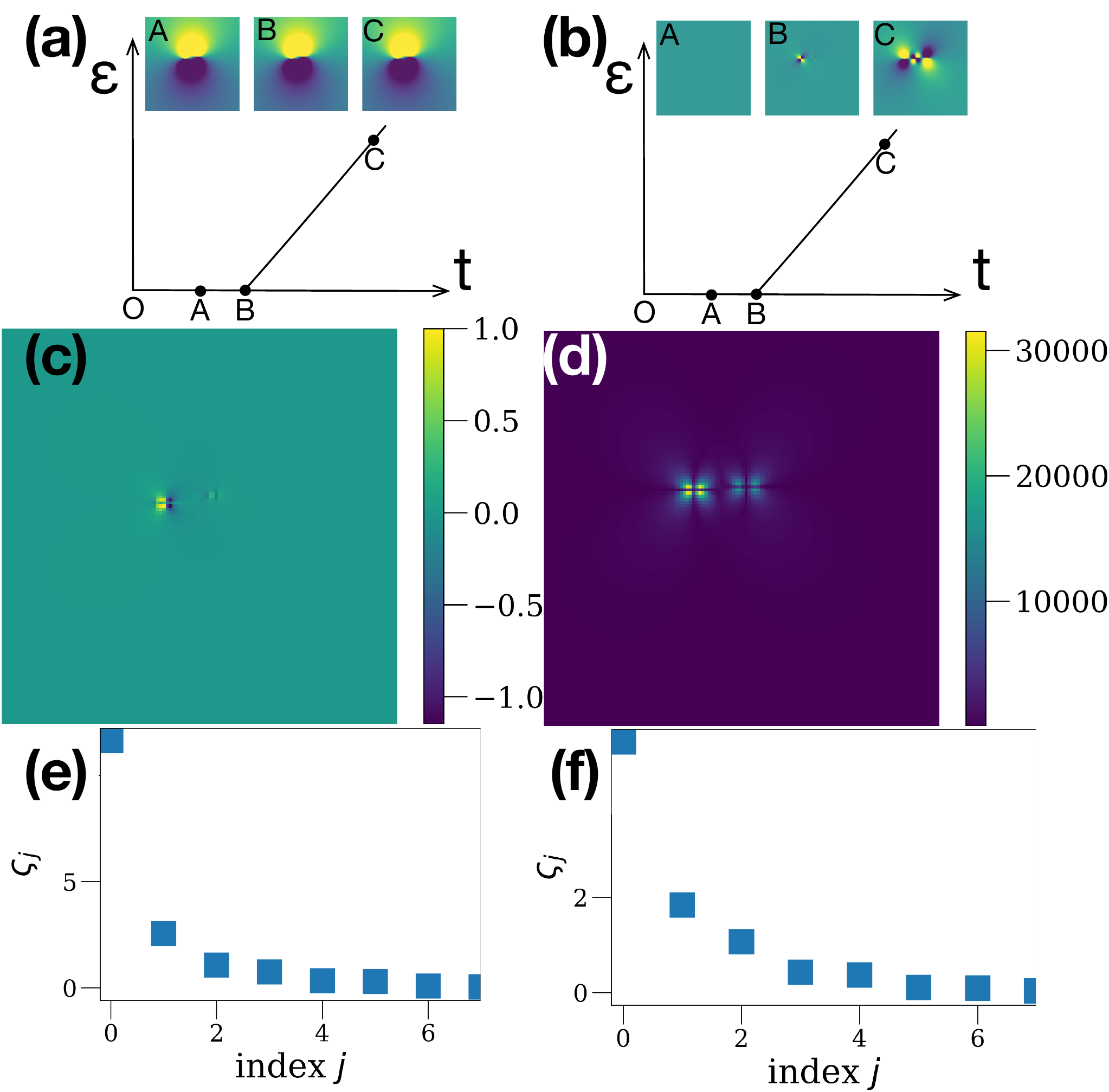}
\caption{{\bf Microstructural Fingerprinting in a Toy Model of Plasticity}: (a) Strain $\veps(t)$ follows a non-linear evolution, controlled by the increase of stress. Two same-sign edge dislocations are pinned at respective obstacles in A, are depinned in B, and glide along their slip-plane along the $x$-direction in C. Insets A, B, C display the corresponding strain invariant images discussed in text. Hyper-reduction leads to well defined modes, the most dominant of which is shown in (c). This solution is analytically confirmed. Analogously, (b, d, f) correspond to the case of dislocation pair nucleation at a specific location (opposite-signed dislocations). Notice the characteristic difference between the calculated modes. The differences are explained in the text.}
\label{fig:2}
\end{figure}

\begin{figure*}[tb]
\centering
\includegraphics[width=0.99\textwidth]{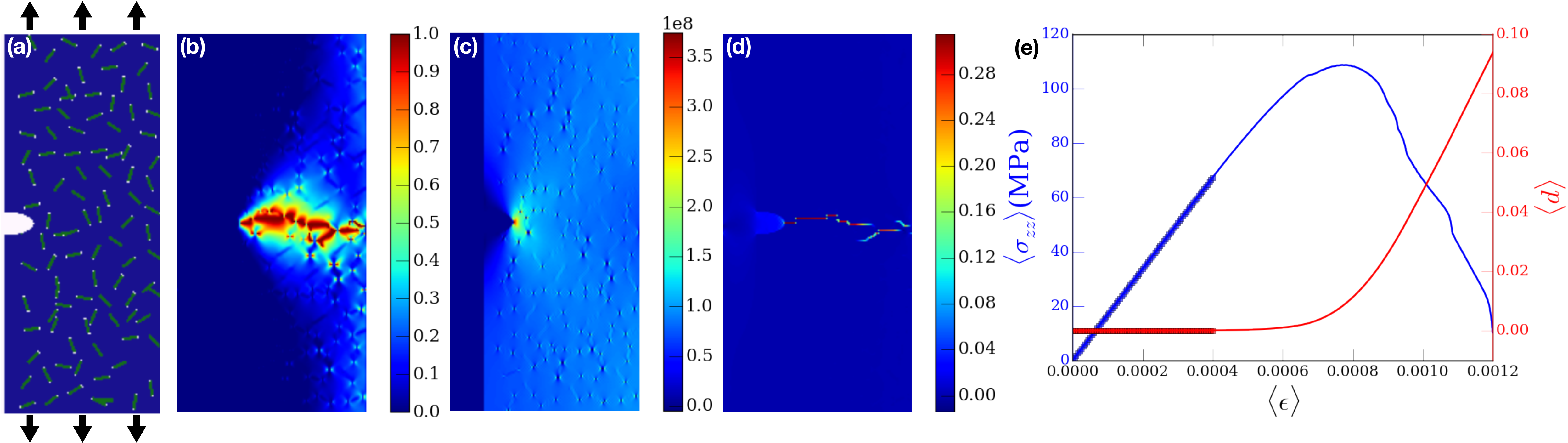}
\caption{{\bf Model binary alloy demonstration \& modeling}: Two-phase material of FCC crystalline structure, that can deform plastically and has brittle fracture characteristics.  (a) Texture of notched specimen with $30$ inclusions of length $240\mu$ m. Arrows point the loading direction. (b) Phase field of damage $\langle d\rangle$ across the sample at a late deformation stage, (c) Uniaxial stress snapshot along the loading axis at an early deformation stage, (d) Uniaxial strain along the loading $z$ axis at a late deformation stage (same as in (b)), (e) Evolution of stress and damage field {\it vs.} imposed loading strain. Thick lines indicate the testing strain (see text).}
\label{fig:3}
\end{figure*}

\subsection{Microstructural fingerprints for a toy model of dislocation dynamics}
\label{sec:C}

In order to demonstrate how microstructural fingerprinting applies in crystal plasticity, we consider a toy example of edge dislocation dynamics. We consider the case of a single slip system, with cores along the $z$-axis and the Burgers' vector in the positive $x$-direction, $\vec b=b \hat x$ with dislocations solely gliding under shear stress along the $x$-direction~\cite{Amodeo:1990aa,groma2003spatial,zaiser2001statistical,van1995discrete}. Here, we  consider a periodic system of size $N\times N$ ($N=40$b), with only two dislocations, which may either be i) pinned at obstacles, i) glide in the same direction or ii) nucleate as a positive-negative dislocation pair at a location $\bfr_0=(x_0,y_0)$. This model is exactly solvable, and provides a way to understand the basic character of the calculations we are considering.

Dislocation motion is assumed to be overdamped, so equation of motion of discrete dislocations can be written as:
\begin{equation}
\dot{x}_i(t) = s_i \left[ \tau_\text{ext} + \sum_{j=1, j\ne i}^N s_j \tau_\text{ind}(\bm r_i -\bm r_j) \right]; \ \dot{y}_i(t) = 0,
\label{eq:eq_mot}
\end{equation}
where $\tau_\text{ext}$ is the externally applied shear stress and $\tau_\text{ind}$ denotes the stress field of an individual positive ($s_i=+1$) dislocation. The latter is calculated for periodic systems by considering an infinite amount of image dislocations both in the $x$ and $y$ directions:
\begin{equation}
\tau_\text{ind}(x,y) = \sum_{i,j=-\infty}^\infty \tau_\text{ind}^\text{ibc}(x-iN,y-jN),
\end{equation}
where
\begin{equation}
\tau_\text{ind}^\text{ibc}(x,y) = \frac{x(x^2-y^2)}{(x^2+y^2)^2}
\end{equation}
is the solution for infinite boundary conditions~\cite{Hirth82}. The equation of motion (\ref{eq:eq_mot}) is solved by a 4.5th order Runge-Kutta scheme.

Analytically, for an independent dislocation at (0,0), and given $D=Gb/(2\pi(1-\nu))$ one has,
$\si_{xx}=-D y \frac{3x^2 + y^2}{(x^2 + y^2)^2}$, 
$\si_{yy}= D y \frac{x^2 - y^2}{(x^2 + y^2)^2}$,
$\si_{zz}=\nu \cdot (\si_{xx}+\si_{yy})$,

and the dislocation pressure is:
\bea
p=\frac{\si_{xx}+\si_{yy}+\si_{zz}}{3} = \frac{2(1+\nu)D}{3}\frac{y}{x^2+y^2}=\nonumber\\  =\frac{(1+\nu)Gb}{3\pi(1-\nu)}\frac{y}{x^2+y^2}
\eea

If one tracks the strain components of this 2D system, then the first strain invariant of an edge dislocation is just,
$I_{\eps}(\bfr) = \sum_{i} \eps_{ii}(\bfr)$. Since $\eps_{zz}=0$ in a 2D system dominated by edge dislocations, then we can use the fact that for isotropic solids we have $\sigma_{zz} = 2G\eps_{zz} + \lambda \sum_{i}\eps_{ii}$ (with $\lambda=\frac{2}{1-2\nu}G$), thus giving:
\bea
I^{+}_{\eps}(\bfr)=\frac{1}{\lambda} \sigma_{zz} = - b \frac{(1-2\nu)\nu}{2\pi(1-\nu)}\frac{y}{x^2+y^2}
\eea
where $+$ denotes the result for a positive dislocation ($\vec b=b\hat x$). 

For a dislocation pinned at an obstacle that starts gliding after the external stress increases beyond a threshold $\sigma_{\rm thr}$ in a non-linear, but differentiable manner (so that $\eps(t)=f(t)$), it is straightforward to estimate the evolution of the strain invariant $I^{\pm}_\eps(\bfr)$:
\bea
\frac{dI^{\pm}_\eps(\bfr)}{dt} = \mp\frac{2f'(t)(x-x_0+f(t))}{(y-y_0)^2 + (x-x_0 + f(t))^2} I_{\eps}(\bfr)
\label{eq:edge}
\eea
Given the simplicity of the problem, the $\tilde\bfJ$ operator is diagonal, thus it is straightforward to infer its properties. It is worth noting that it is non-zero only when dislocations are in motion, and it has a characteristic left-right asymmetry, with respect to the original pinning point location. For a negative dislocation~\cite{Hirth82}, from Eq.~\ref{eq:edge}, $\tilde\bfJ$ is exactly negative, leading to a very drastic difference in the fingerprint of a nucleating dislocation pair compared to a pair of gliding dislocations (\cf Fig.~\ref{fig:2}).

SEA manifests its usefulness in identifying distinguishable signatures of characteristic events (dislocation glide {\it vs.} pair nucleation) through a parameter-free, experimentally tractable and automatic manner.
}

\begin{figure*}[tbh]
\centering
\includegraphics[width=0.95\textwidth]{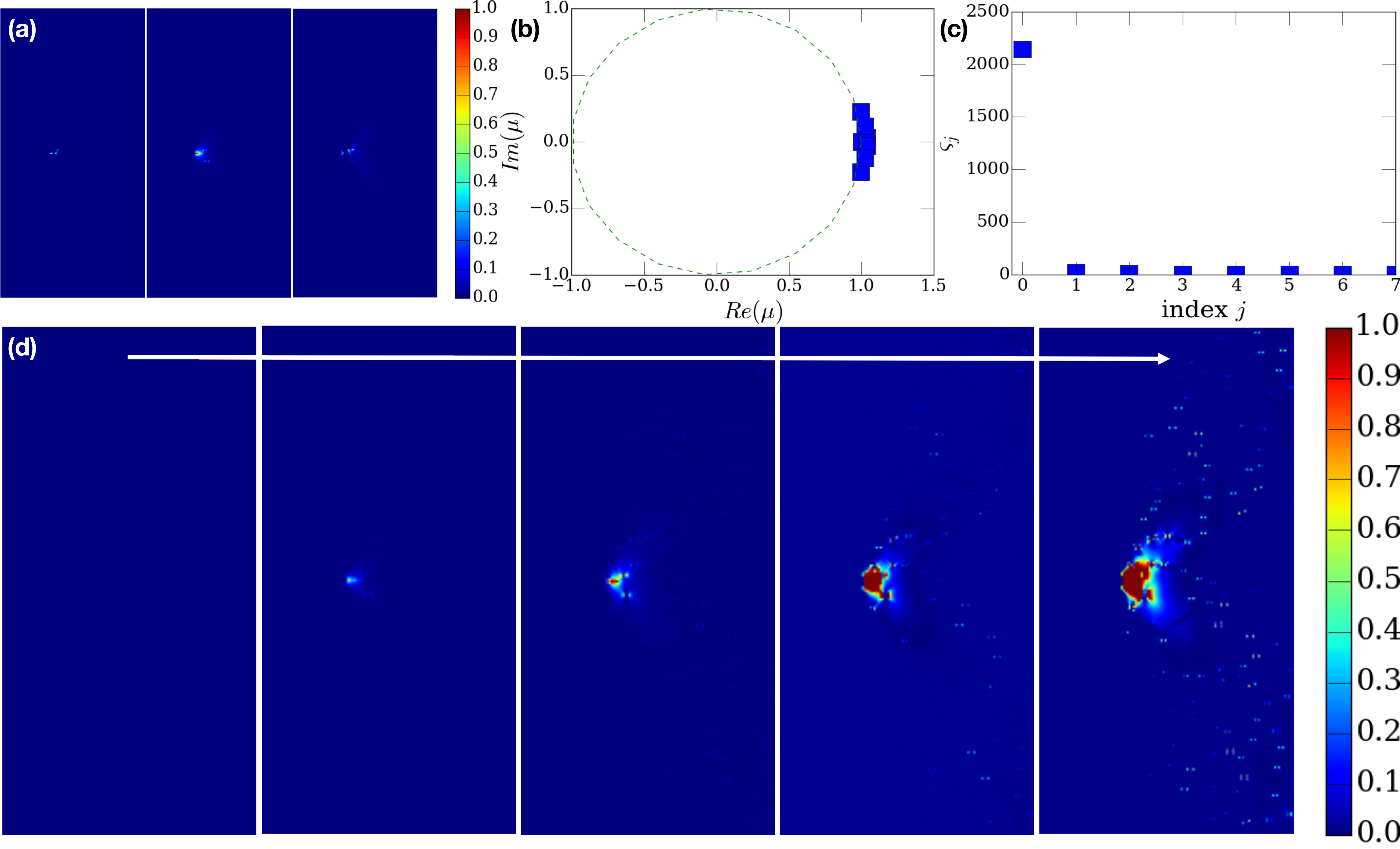}
\caption{{\bf Demonstration of Single-Sample Mode (SSM) Identification \& Prediction using $\phi\equiv 1-d$}. (a) Selected (3) EIMs emerging from considering the consecutive damage images until close to failure. (b) real and imaginary parts of $\ln\lambda_k$ for each kept k-mode. (c) Singular value $\varsigma_j$ for the j-th SVD-mode. (d) Predicted damage evolution snapshots using only the identified modes.}
\label{fig:4}
\end{figure*}
\begin{figure}[t]
\centering
\includegraphics[width=0.5\textwidth]{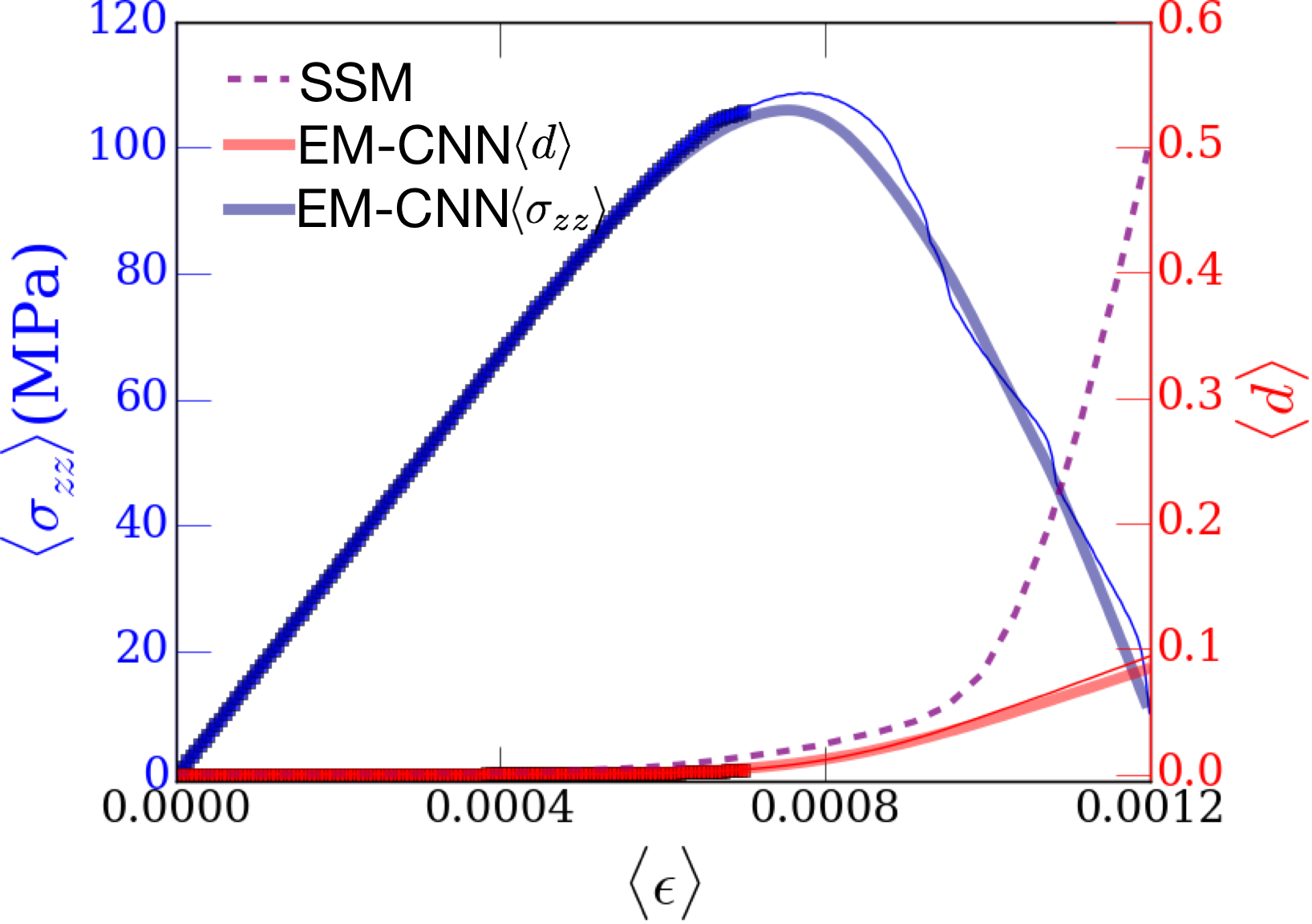}
\caption{{\bf Equation-Free Predictions with $\phi\equiv 1-d$:} The SSM prediction extends only to the behavior of the damage field. EM-CNN uses the library of prior samples and modes $\libPsi$ to reconstruct a predicted behavior that is very accurate for the currently modeled stochastic microstructures.}
\label{fig:5}
\end{figure}

\begin{figure*}[tbh]
\centering
\includegraphics[width=0.95\textwidth]{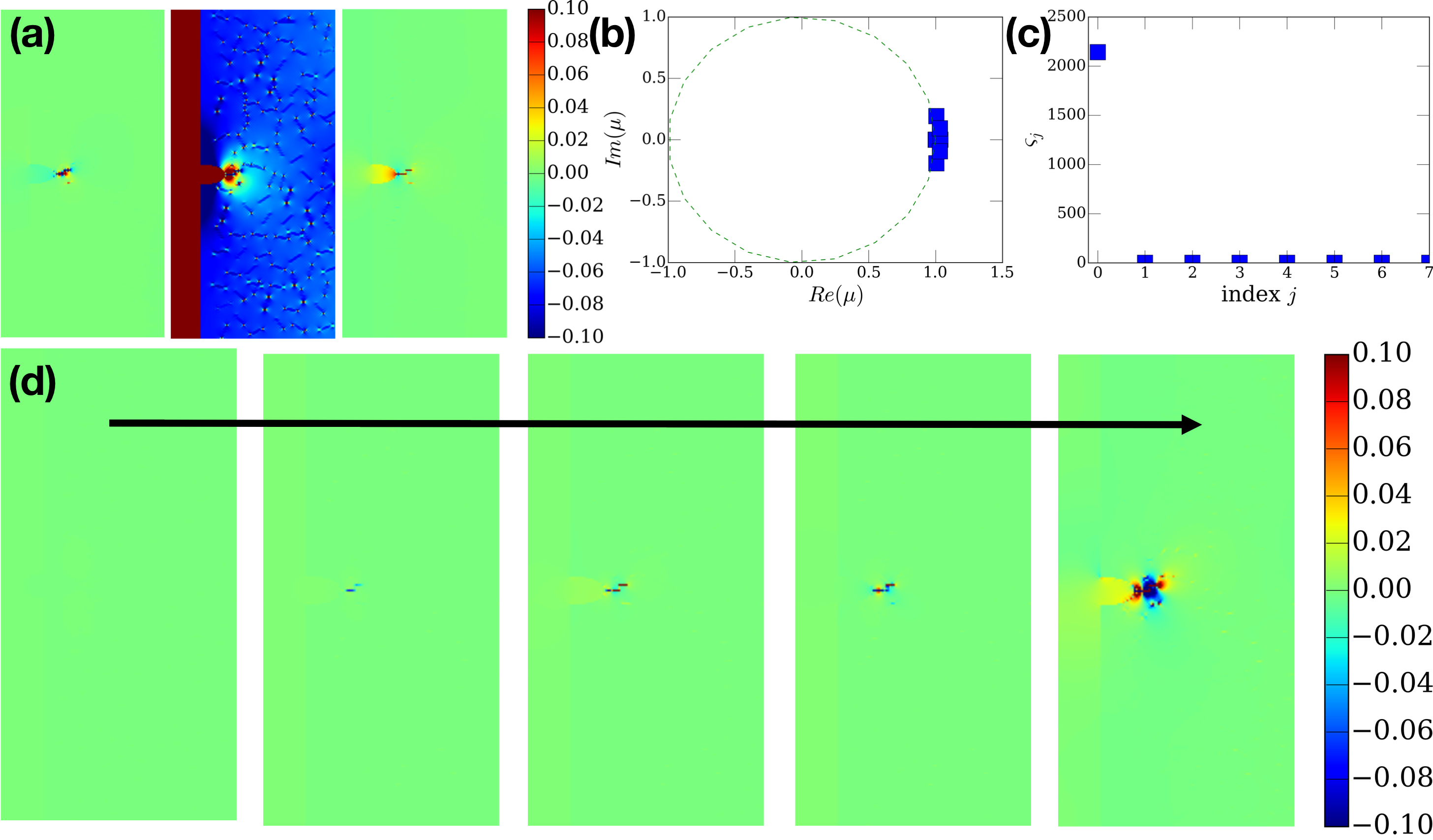}
\caption{{\bf Demonstration of Single-Sample Mode (SSM) Identification \& Prediction using $\phi\equiv I_{1}^{(\epsilon)}$}. (a) Selected (3) EIMs emerging from considering the consecutive damage images until close to failure. (b) real and imaginary parts of $\ln\lambda_k$ for each kept k-mode. (c) Singular value $\varsigma_j$ for the j-th SVD-mode. (d) Predicted damage evolution snapshots using only the identified modes.}
\label{fig:6}
\end{figure*}

\begin{figure}[tbh]
\centering
\includegraphics[width=0.5\textwidth]{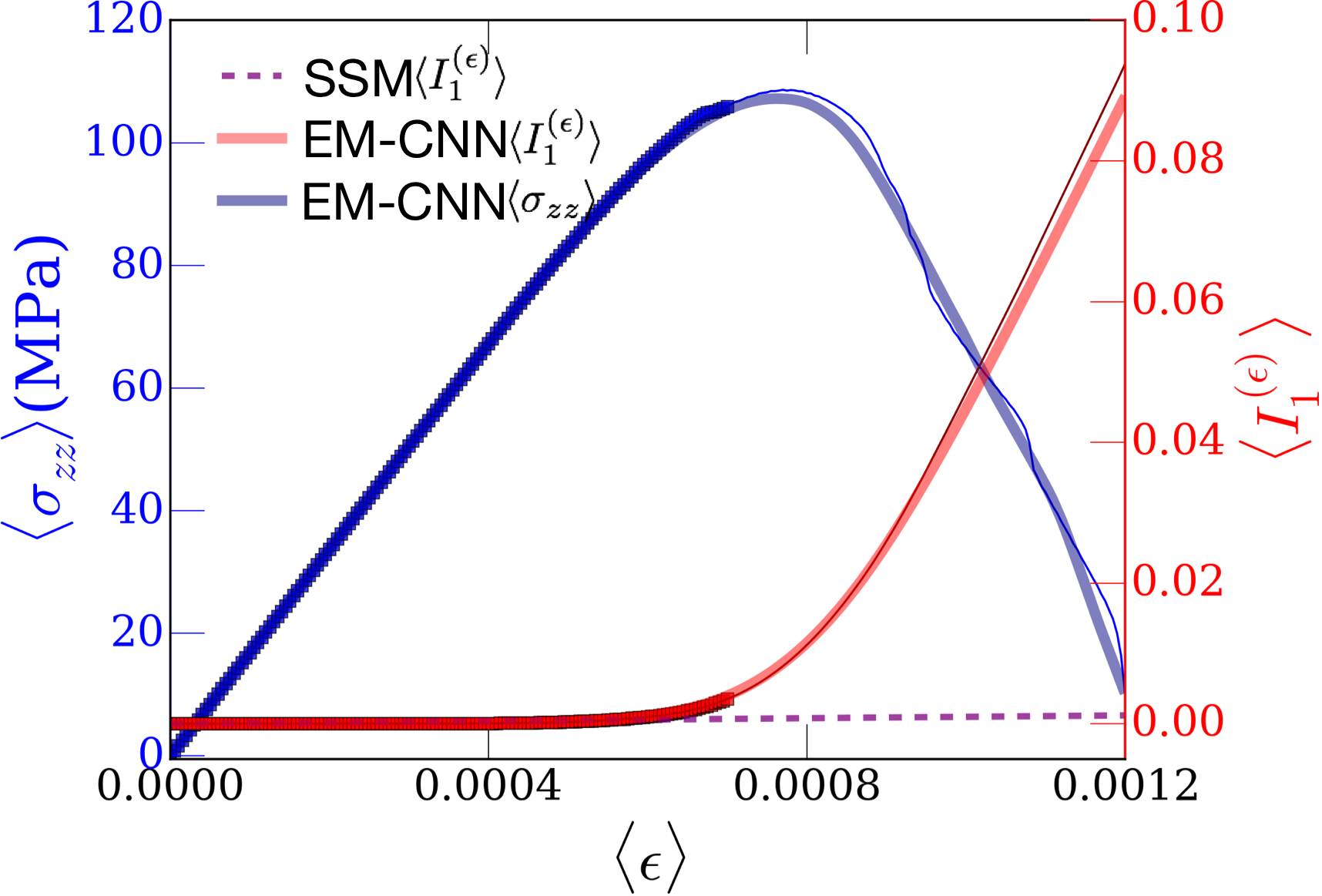}
\caption{{\bf Equation-Free Predictions with $\phi\equiv I_{1}^{(\epsilon)}$:} The SSM prediction extends only to the behavior of the damage field. EM-CNN uses the library of prior samples and modes $\libPsi$ to reconstruct a predicted behavior that is very accurate for the currently modeled stochastic microstructures.}
\label{fig:7}
\end{figure}

\subsection{Microstructural Fingerprints and Data-Rich Predictions: Plasticity and Damage in a Model Binary Alloy}
\label{sec:D}

While toy examples are insightful in showing the origin and potential usefulness of SEA, it is natural to inquire its applicability in realistic situations where plasticity,  damage, as well as fracture are plausible instabilities. In such complex cases, there are strong spatial and temporal correlations that make analytical calculations of the $\bfJ$ matrix impossible. However, SEA may efficiently estimate the predominant EIMs in numerical fashion. 

To explicitly illustrate the method, we  consider an exemplary test case~\cite{pap2019scg,Bazant:1997hx, Asaro:2006fr,Roters:2019be,Asaro:2006fr} by laterally loading a notched thin-film specimen in contact with air in the horizontal direction, with sample dimensions: $0.25cm^2$ and a resolution at $40\mu$m in the loading direction, $20\mu$m the horizontal, and $40\mu$m in thickness -- RVE's size (20$\times$40$\times$40$\mu$m$^3$)).The notch facilitates crack growth and it has width $0.15$mm and height $0.3$mm, with an elliptical shape. The crystalline structure of the matrix material is FCC, with elastic coefficients of the material are $C_{zz}=150$GPa, $C_{zy}=120$GPa, $C_{xx}=80$GPa (with $x$ being the film thickness direction). The importance of this example is that its dimensions can be efficiently achieved by current experimentation procedures~\cite{Schreier:2009xd}. The sample also contains needle-like inclusions with width $80\mu$m and fixed length that could be either 80, 160, 240, 320 $\mu$m with FCC crystalline structure but distinct material properties ($C^{incl}_{zz}=180$GPa) from the matrix and there is also imposed microscale damage at the inclusion tips. The inclusions are placed randomly in the sample and their number can be either 0, 2, 5, 10, 15, 20, 25, 30, leading to total $32$ possible microstructures (a texture is shown in Fig.~\ref{fig:3}(a)).  Cases of different microstructures can be thought of as binary alloys with different aging conditions or compositions. Cases of different initial realizations can be thought of different samples at the same nominal composition and processing history.

We utilize a phase field model~\cite{Ambati:2015bi} in the continuum to solve for material deformation due to elasticity, plasticity and damage evolution within the sample. Details of the model's hardening and damage dynamics can be found in Refs.~\cite{pap2019scg,Roters:2019be}. In summary, the model captures finite deformations in a cubic grid, which are used to calculate constitutively plastic distortion rates along all 12 FCC slip systems, as well as damage evolution. The model is solved using a spectral approach which promotes numerical stability for highly disordered microstructures~\cite{pap2019scg}. As shown characteristically in Fig.~\ref{fig:3}, the model predicts the damage of the sample due to a noisy crack that emanates from the notch and the sharp inclusion tips.  In the model, fracture takes place at a loading stress $\sim 100$MPa, controlled by both Linear Elastic Fracture Mechanics~\cite{Sanford:2003oc}, as well as quasi-brittle fracture induced by the inclusions~\cite{Bazant:1997hx,pap2019scg}. 
 Damage/stress/strain profiles generated in our simulations (Fig.~\ref{fig:3}(b-d)) can be easily resembled to profiles that may be generated by Digital Image Correlation (DIC) techniques~\cite{Schreier:2009xd}.
For the implementation of SEA, we consider as $\phi$ fields, the phase-field damage field $d$ and the first strain invariant $I^{\eps}_1\equiv \eps_{xx}+\eps_{yy}+\eps_{zz}$. An important parameter is the testing strain $\eps_t$, which is set at $0.04\%$ strain in the example of Fig.~\ref{fig:3} but it is varied from 0 to 0.07$\%$.

If SEA is applied on the data shown in Fig.~\ref{fig:3}, then EIMs can be estimated. The results for $\phi\equiv d$ are shown in Figs.~\ref{fig:4}(a), while the results for $\phi\equiv I_{1}^{\eps}$ are shown in Figs.~\ref{fig:6}(a), using $\eps_{t}=0.07\%$ and $30$ inclusions of length $240\mu$m. As one can see in Figs.~\ref{fig:4}(b), \ref{fig:6}(b), the modes have eigenvalues $\mu$ that denote positive LEs ($Re(\mu)>1$), while they also have an oscillating component. Characteristically, though, the strain deformation is primarily dominated by one mode (\cf Figs.~\ref{fig:4}(c), \ref{fig:6}(c)) which essentially corresponds to the damage instability at the notch location.

By using the EIMs' information, one may try to partially reconstruct the damage evolution in the sample (\cf Fig.~\ref{fig:4}(d)) or the strain-invariant evolution (\cf Fig.~\ref{fig:6}(d)). This reconstruction consists of the Single Sample Mode (SSM) prediction (\cf Eq.~\ref{eq:pred}), that is promoted by considering the formal mode expansion into modes and then extrapolating the modes into the future, as in any non-linear dynamical system~\cite{Strogatz:1994cu,guck}. The SSM is able to capture the incipient instability, even though equation-free predictions of average quantities typically miss the details of the true response (for example, compare Fig.~\ref{fig:4}(d) with Fig.~\ref{fig:3}(b)). This is naturally expected for investigations of precursors in bifurcation dynamics~\cite{Papanikolaou:2017zl}. Nevertheless, it is quite promising that the 8 captured EIMs can predict the onset of damage and provide a simplified damage evolution. This is a signature of EIMs being appropriate microstructural fingerprints of the structural response.

The main idea behind microstructural fingerprinting is that the EIMs (irrespective of which $\phi$-field they are based on) may be directly compared to any superposition of other modes $\bfPsi_i$ that have been similarly calculated from other microstructures. The EIMs $\bfPsi_i$ are stored in the library ${\rm lib}_\bfPsi$, together with all tested mechanical responses for the sample in question ($\phi_i=\mathcal{H}_i(\epsilon)$) (\cf Fig.~\ref{fig:plan}). In principle, eigen-decomposition of $\Psi_0$ allows for the existence of probability weights $w_i$ that should satisfy: $\bfPsi_0 = \sum w_i \bfPsi_i$~\cite{Press:1987oq}. Then, we conjecture that if all other pre-existing library samples have been tested to failure, then we also have that,
\bea
\langle\phi_0\rangle = \sum_{i=1}^{N_s} w_i \mathcal{H}_i(\eps)
\eea

Namely, various different microstructures, tested at total strain $\eps_0\leq 2\%$ with identical boundary conditions may {\it exactly} model the mechanical response of an {\it unknown} microstructure, using appropriately weighted sums. For making accurate, data-rich predictions of damage and plasticity we utilize a library of EIMs $\bfPsi_i$ and complete responses up to failure $\mathcal{H}_i(\eps)$. Major understanding of elastic instabilities originates in the fundamental works by Eshelby~\cite{Eshelby:1957bd}. Here, we propose a systematic capture of a scalable set of defects in a consistently tracked manner that extend those studies and understanding.
Thus, it is assumed that $r$ modes $\bfPsi_i\equiv\{\psi_j\}$ for $j\in[1,r]$ are precisely known for sample $i$ that were previously tested up to complete failure for desired loading conditions, each providing a functional form $\phi_i=\mathcal{H}_i(\epsilon)$ where $\phi_i$ denotes the loading response of interest ($\sigma$, $\tau$, etc.) while $\epsilon$ provides the loading probe of interest (\eg loading strain in a particular direction). 

\subsection{Deep Convolutional Neural Networks and General Framework For Using Microstructural Fingerprinting Towards Mechanical Predictions Up to Failure}
\label{sec:E}

The numerically approximate character of SEA enforces the use of advanced approaches that may efficiently compare, classify and reconstruct fingerprints. The data sets $\{\bfPsi,\mathcal{H}\}$ are produced for different microstructures but same loading and boundary conditions for precise comparison purposes, so that weights are identified that provide the following equality,
\bea
\bfPsi_{0}^{(unknown)} \simeq \sum_{i\in {\rm lib}_\bfPsi} w_i \bfPsi_{i}
\eea

The identification of probability weights $w_i$ requires a projection of the library EIMs on the new EIM $\bfPsi_0$. For estimating the probability weights $w_i$, we utilize a dCNN. The implementation is straight-forward in that mode identification is treated as a face-recognition problem. While there are multiple approaches towards identifying appropriate mode projections on the emerging basis~\cite{Cariolaro:2015ho}, we find that d-CNNs are efficient and robust. CNNs were originally suggested for handling two dimensional inputs (e.g. image), in which features learning were achieved by stacking convolutional layers and pooling layers.\cite{LeCun:1998xz} dCNNs (dCNN) were introduced~\cite{dnn}, for improving performance in image recognition. CNNs and dCNNs are well fit for automated defect identification in surface integration inspection, and their optimization is based on backpropagation and stochastic gradient descent algorithms.~\cite{def1,def2,def3,def4} We apply a standard deep convolutional neural net for image recognition of similar resolution to our images, by using the TensorFlow software~\cite{tf}. The predictions of this Elastic Mode Convolutional Neural Network (EM-CNN) approach for the average  field $\phi$ (either damage field $\dav$ or first strain invariant $I_{1}^{\eps}$) and loading stress field average $\sav$ are shown in Figs.~\ref{fig:5} and ~\ref{fig:7}, whereas true mechanical response (\cf Fig.~\ref{fig:3}(e)) is overlayed. The EM-CNN results are efficient and robust, agreeing with the true mechanical response. The success of the method is clearly connected to the fact that the library of pre-existing data ${\rm lib}_\bfPsi$ contained similar microstructures to the ones tested. However, we wish to emphasize that the process is fundamentally not interpolation, but instead an identification of proper statistical averaging.

\begin{figure}[tb]
\centering
\includegraphics[width=0.5\textwidth]{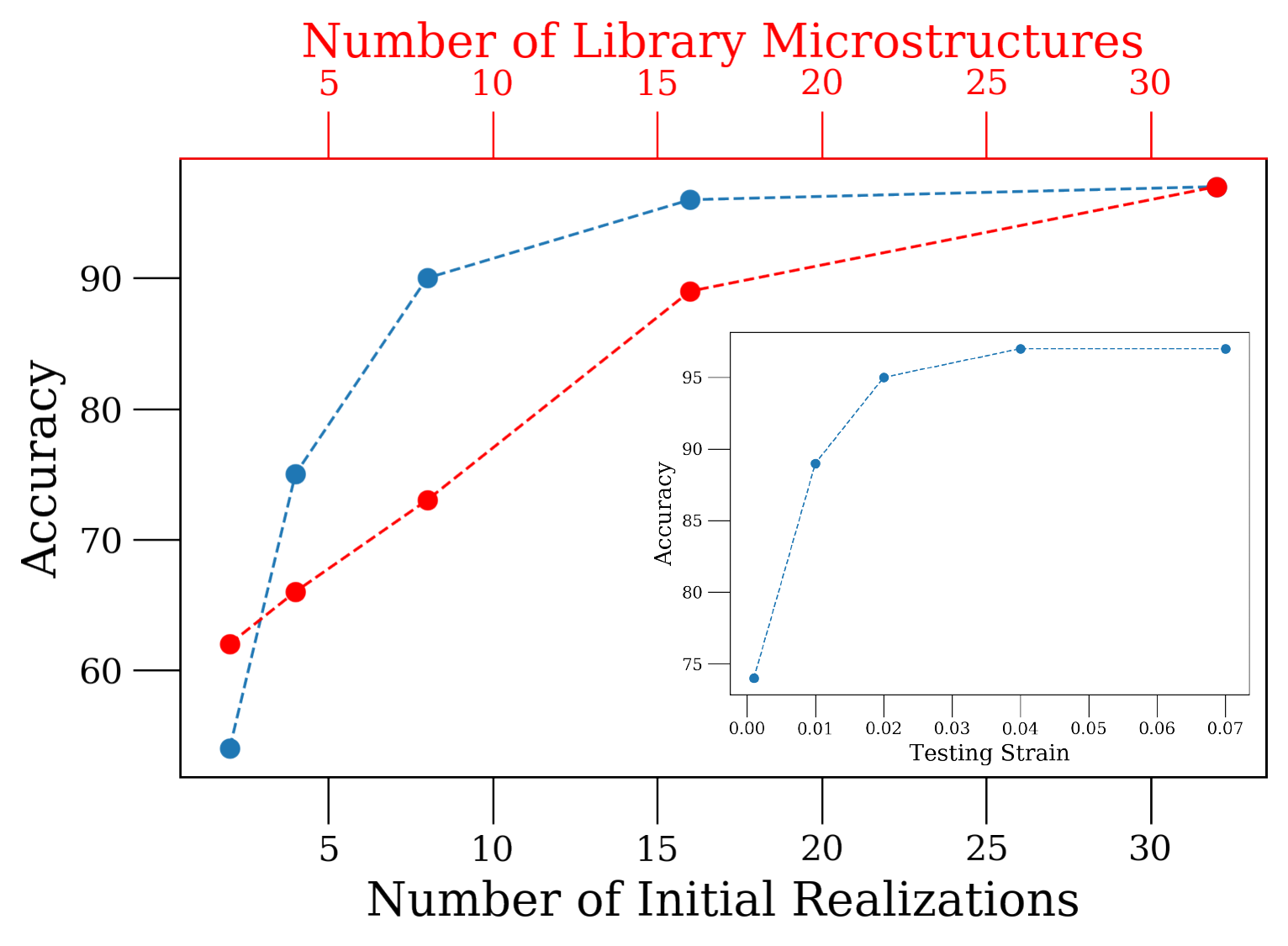}
\caption{{\bf Accuracy:} Effect of the number of initial conditions and the number of possible microstructures (various inclusion lengths) used in the synthetic data for producing data such as in Fig.ccc. The accuracy quickly grows beyond 90\% for $>10$ initial realizations for each of $>10$ available library microstructures. The inset shows the effect of the testing strain, displaying that the accuracy is beyond 90\% for testing strain $>0.1\%$ }
\label{fig:8}
\end{figure}

The interpretation of the weights $w_i$ should be made in terms of the identification of an appropriate combined set of defects that provide similar modes to the sample of interest. 
Testing a large variety of microstructures in advance, may allow for the identification of EIMs for a completely unknown microstructure that may include completely different compounds through cooperation of different types of microstructures. 

The success of  SEA for predictions of mechanical properties can be estimated by the accuracy (fraction of correctly classified samples) for different number of available library microstructures (different number/length of inclusions in sample), number of initial random realizations but with same qualitative behavior (same number of inclusions, different random locations/orientations), different small deformation testing strain (where EIMs are calculated). The results are summarized in Fig.~\ref{fig:8}. The behavior consistently points to perfect predictions when the number of available microstructures is larger than 15 and the testing strain is larger than 0.02\%. In this work, we did not focus on optimizing the performance of the dCNN used~\cite{tf}, a topic that will be the focus of forthcoming publications.

\subsection{Conclusions}
\label{sec:F}

In this work, we demonstrated a parameter-free approach, labeled as SEA, that is focused on characterizing the stability of elasticity, towards understanding and predicting mechanical properties of solids that may deform towards plasticity or/and damage. The main outcome of this analysis are   dynamical modes (EIMs) that characterize spatial instabilities of elasticity. EIMs can be considered as microstructural fingerprints. We presented the theory of the approach and then we demonstrated it in a toy example of crystal plasticity that provided analytical insights for EIMs' origin and character. Using the developed methodology, we applied SEA to a realistic model of plasticity and damage for a model binary alloy. Through this investigation, and with help from dCNNs, we showed that the use of EIMs as fingerprints can lead to successful mechanical predictions for microstructures that are only tested at small deformations.

The usefulness of SEA can be either towards a supplementation multiscale materials modeling or for producing data-rich predictions of mechanical properties in untested microstructures. Multiscale materials modeling requires a variety of signatures/tests that may provide verifiable links across scales towards modeling accuracy; EIMs may provide such signatures. In addition, SEA's implementation may be solely focused on experimental settings, where microscopy may contribute spatially resolved information that may be used towards identifying and classifying EIMs. Future studies will explore SEA's efficacy in both experimental and modeling fronts.

\begin{acknowledgements}
This work is supported by the National Science Foundation, DMR-MPS, Award No \#1709568.
\end{acknowledgements}


%

\end{document}